\documentclass[apjl]{emulateapj}

\def\ltsim{~\rlap{\lower -0.5ex\hbox{$<$}}{\lower 0.5ex\hbox{$\sim\,$}}}

\slugcomment{Published in The Astrophysical Journal, 683, L103-L106, 2008 August 20}

\shorttitle{OUTER-DISK PROPERTIES OUTSIDE BREAK RADIUS}
\shortauthors{BAKOS, TRUJILLO, \& POHLEN}

\begin{document}


\title{COLOR PROFILES OF SPIRAL GALAXIES: CLUES ON OUTER-DISK FORMATION SCENARIOS}


\author{Judit Bakos}
\author{Ignacio Trujillo\altaffilmark{1}}
\affil{Instituto de Astrof\'isica de Canarias, E-38205, La Laguna, Tenerife, Spain; jbakos@iac.es, trujillo@iac.es}
\and
\author{Michael Pohlen}
\affil{Cardiff University, School of Physics \& Astronomy, Cardiff, CF24 3AA, Wales, UK; Michael.Pohlen@astro.cf.ac.uk}


\altaffiltext{1}{Ram\'on y Cajal Fellow}


\begin{abstract}

We have explored radial color and stellar surface mass density profiles for a
sample of 85 late-type spiral galaxies with deep (down to $\sim$27
mag arcsec$^{-2}$) SDSS $g'$ and $r'$ band surface brightness profiles. About $90\%$
of the light profiles have been classified as broken exponentials, exhibiting either truncations (Type~II galaxies) or antitruncations (Type~III galaxies). The color profiles of Type~II galaxies show a ``U shape'' with a minimum of $(g' - r') = 0.47\ \pm\ 0.02$ mag at the break radius. Around the break
radius, Type~III galaxies have a plateau region with a color of $(g' - r') =
0.57\ \pm\ 0.02$. Using the color to calculate the stellar surface mass density
profiles reveals a surprising result. The breaks, well established in the light
profiles of the truncated galaxies, are almost gone, and the mass profiles now
resemble those of the pure exponential (Type~I) galaxies. This result
suggests that the origin of the break in Type~II galaxies is more likely due to
a radial change in stellar population than being associated with an actual drop
in the distribution of mass. Type~III galaxies, however, seem to preserve their
shape in the stellar mass density profiles. We find that the stellar surface
mass density at the break for truncated galaxies is $13.6 \pm 1.6 \ {M}_{\sun}$pc$^{-2}$ and for the antitruncated ones is $ 9.9 \pm 1.3 \ {M}_{\sun}$pc$^{-2}$ for
the antitruncated ones. We estimate that the fraction of stellar mass outside
the break radius is $\sim$15$\%$ for truncated galaxies and $ \sim$9$\%$ for
antitruncated galaxies. 

\end{abstract}

\keywords{galaxies: evolution - galaxies: formation - galaxies: photometry - galaxies: spiral - galaxies: structure}



\section{Introduction}
\label{intro}

Our picture regarding the diversity of the radial surface brightness profiles
of spiral galaxies has changed greatly since the early work of Patterson (1940)
and de Vaucouleurs (1958), who showed that the disks of spiral galaxies
generally follow an exponential decrease in their radial surface brightness
profile. Nowadays, this view has become clearly insufficient, as not all the disk galaxies (indeed, only the minority for late-types) are well described
with a single exponential fitting function as shown in several recent studies
(Erwin et al.~2005; Pohlen \& Trujillo 2006, hereafter PT06; Florido et al. 2006, 2007; Erwin et al. 2008), where they have identified three basic classes of surface
brightness profiles depending on an apparent break feature or lack of one: (1)
the pure exponential profiles (Type I) with no breaks, (2) Type II with a
``downbending break'' (revising and extending a previous classification
introduced by Freeman [1970] to include the so-called truncations of the
stellar populations at the edge of the disk discovered by van der Kruit
[1979]) and (3) a completely new class (Type III), also described by a broken
exponential but with an upbending profile. The latter, discovered by Erwin et
al. (2005), is also termed antitruncated.

PT06 explored a sample of nearby late-type galaxies using
the Sloan Digital Sky Survey to create a statistically representative set of
radial surface brightness profiles. They found that about 60$\%$ of the
spirals are truncated (Type~II), 30$\%$ are antitruncated (Type~III), and only
10$\%$ have no detectable breaks (Type~I).

Still, little is known about the nature of the breaks or about the presence of
stars beyond that feature. In such low-density ($\le$ 10 $ {M}_{\sun}$pc$^{-2}$) environments at the galaxy peripheries, current star formation
theories forbid efficient star formation (Kennicutt 1989; Elmegreen \&
Parravano 1994; Schaye 2004). However, UV observations (Gil de Paz et
al. 2005; Thilker et al. 2005) have shown ongoing star formation in these
outer regions. In addition, it is clear that there are large number of stars
in the outskirts of galaxies (see, e.g., PT06). 

Several theories have investigated the formation of breaks in case of the
Type~II morphology. Proposed models to explain the existence of truncations in
stellar disks can be grouped into two branches depending on the
relevant mechanism that causes the break: (a) models related to angular
momentum conservation in the protogalactic cloud (van der Kruit 1987) or
angular momentum cutoff in cooling gas (van den Bosch 2001), and (b) models
that attribute the existence of breaks to star formation thresholds
(Kennicutt 1989; Elmegreen \& Parravano 1994; Schaye 2004). In agreement with
this last scenario, Elmegreen \& Hunter (2006) suggested a multicomponent star
formation that would result in a double exponential surface brightness
profile as observed for Type~II galaxies. Recent developments, some of them
combining pieces of the aforementioned scenarios, conclude that secular
evolution driven by bars or spiral arms or even clump disruptions can result
in truncated exponential profiles (Debattista et al. 2006; Bournaud al. 2007;
Ro\v skar et al. 2007; Foyle et al. 2008). Magnetic fields have also been considered to explain the existence of truncations (Battaner et al. 2002). On the other hand, the Type~III
morphology is proposed to be explained by a tidal stripping within a minor
merger (Pe\~narrubia et al. 2006; Younger et al. 2007), by a bombardment of
the disk with dark matter subhalos (Kazantzidis et al. 2007) or by a high
eccentricity flyby of a satellite galaxy (Younger et al. 2008).

In this Letter we show for the first time color and stellar surface mass
profiles of a large sample of galaxies (the PT06 sample) to quantify the
stellar mass density at the break position and the fraction of stellar mass
beyond the break. These values become important while comparing observations
to the results of numerical simulations for the outer-disk formation. In order
to fully understand the galaxy formation and evolution process, it is necessary to
perform a detailed study of the stellar population properties at the galaxies'
outskirts. This kind of study gives insight into how the star formation progresses in the
different parts of the disks, providing clues on the stellar mass
buildup process.

\section{The Data and Analysis Techniques}\label{data}

Our data are the 85 SDSS $g'$ and $r'$ band surface brightness profiles  
 published in PT06. The galaxies were selected to be a representative, volume
limited ($R \ltsim 46$ Mpc) sample of face-on to intermediate-inclined
late-type disk galaxies brighter than ${M}_B = -18.4 $ mag. In that
sense, they range from fainter to brighter surface brightness, from
lower to higher mass, and also from smaller to larger size.
The surface brightness profiles are classified as 9 Type~I, 39 Type~II
and 21 Type~III, i.e., exponential, truncated and antitruncated profiles,
respectively (see PT06 and Erwin et al. 2008 for more details).

The surface brightness limits on our individual surface brightness profiles
(27.0 and 27.5 mag arcsec$^{-2}$ for $r'$- and $g'$- bands, respectively) were estimated by computing when either over- or undersubstracting the sky by $\pm$1$\sigma$ has an effect of more than 0.2 mag
on the surface brightness distribution. These limits were established using two
different methods for determining the sky. In addition, by comparing our
profiles with deeper data (when available), we did not detect any systematic
error in the sky determination. It is worth noting that, for the work presented
here where we combine several profiles to explore the mean properties of the
surface brightness profiles, the uncertainties in the mean properties due to sky
substraction uncertainties are reduced to $\lesssim$0.03--0.04 mag in the outermost
regions of the galaxies.

\begin{figure}[t]
\centering
\includegraphics[scale=1.0]{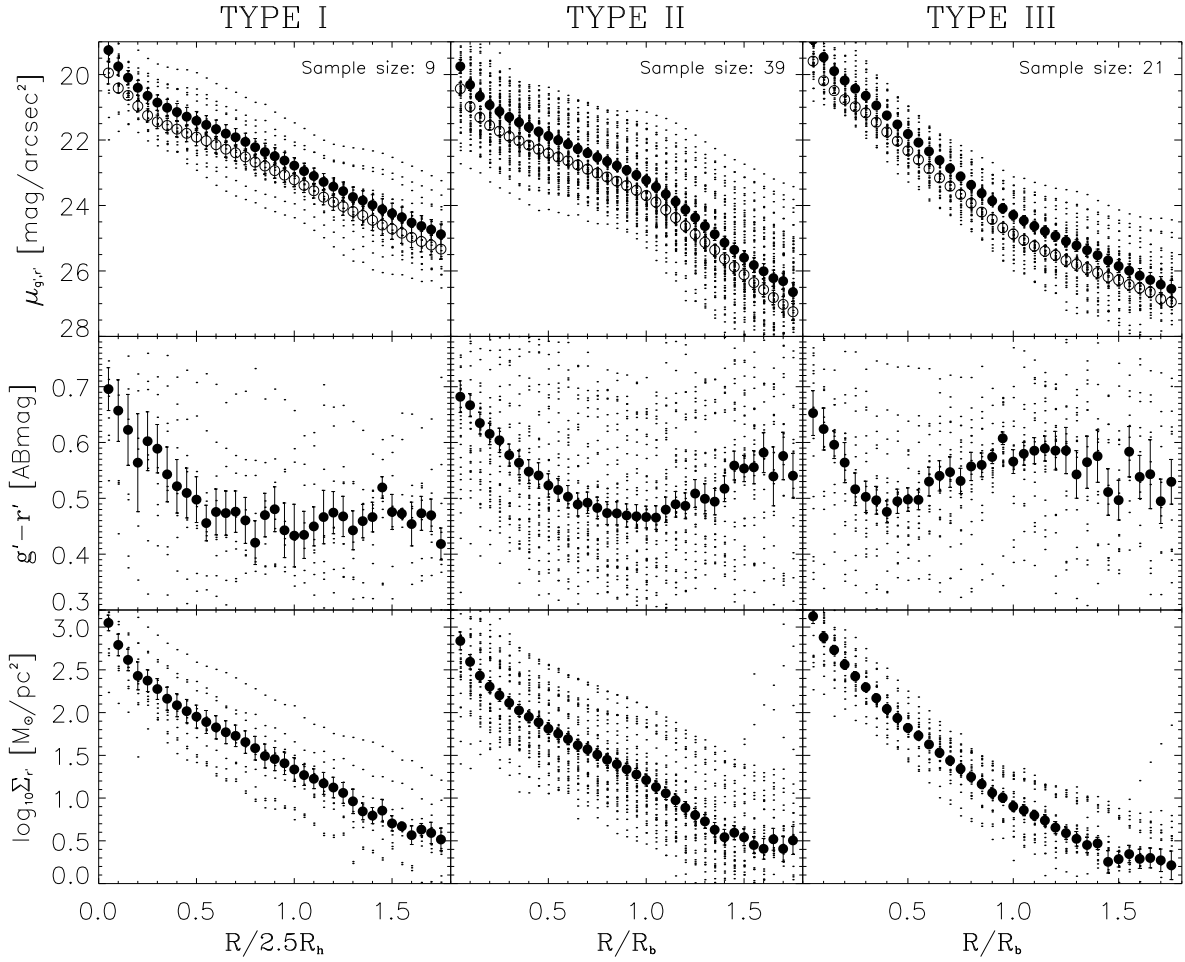}

\caption{\textit{Top row}: Averaged, scaled radial surface brightness
  profiles of 9 Type~I, 39 Type~II and 21 Type~III galaxies. The filled circles
  correspond to the $r'$-band mean surface brightness, the open circles to the
  mean $g'$-band data. The small dots are the individual galaxy profiles in both
  bands. The surface brightness is corrected for Galactic extinction. 
  \textit{Middle row}: $(g' - r')$ color gradients. The averaged profile of
  Type~I reaches an asymptotic color value of $\sim$0.46 mag being
  rather constant outward. Type~II profiles have a minimum color of $0.47 \pm
  0.02$ mag at the break position. The Type~III mean color profile has a redder value of about $0.57 \pm 0.02$ mag at the break. \textit{Bottom row}: $r'$-band surface mass density profiles obtained using the color-to-M/L conversion of Bell et al. (2003). Note how the significance of
  the break almost disappears for the Type~II case. The error bars are given as
  $\sim\sigma/\sqrt{N}$, where $\sigma$ is the scatter and $N$ is the number
  of galaxies taken into account for estimating the mean averaged value in each
  bin. These error bars do not account for uncertainties in the sky
  determination.}

\label{colorgrads}
\end{figure}

We have removed 16 galaxies from our original dataset due to peculiarities of
the classification, for example, all of the Type~II-AB galaxies where the
apparent break is to some extent artificial (see detailed discussion in PT06). We also
excluded IC 1067, which has a very uncertain classification, being either
Type~II or a possible Type~I. For galaxies with mixed classifications (see
PT06) we only used the first type. 
To statistically compare the surface brightness profiles of all galaxies in
our sample, we normalized the sizes of the different galaxies according to
their respective $r'$-band break radii (see Fig.~\ref{colorgrads}).
For the Type~I galaxies, lacking a break radius, we applied 2.5 times the
measured scale length as a normalizing factor, which is the typical radius of
the break for the Type~II galaxies (PT06).
The Type~II and Type~III galaxies have their $r'$-band 27 mag arcsec$^-2$ isophote at around 1.8 times the break radius, which constrains how far we can accurately trace out the behavior of the light profiles into the outskirts of the disks.

In order to calculate a robust mean color that characterizes our sample, we
removed the color of each individual galaxy at a given radii where the $g'$- or
the $r'$-band surface brightness is below the above critical limits at that
radii. We have obtained a robust mean value of the color for all galaxies by
removing the $3\sigma$ outliers. We have explicitly checked that using a cut in
the surface brightness does not result in a bias towards any absolute magnitude
range.

The results of this mean are the profiles shown in the middle row of
Figure \ref{colorgrads}. It is straightforward to link the stellar mass density
($\Sigma$) profile with the surface brightness profile at a given wavelength
($\mu_{\lambda}$) if we know the mass-to-light ($M/L$) ratio, using the expression below.
\begin{equation}
 log_{10} \Sigma = log_{10} (M/L)_{\lambda} - 0.4(\mu_{\lambda} - m_{abs,\sun,\lambda}) + 8.629,
 \label{sigma}
\end{equation}
where $m_{abs,\sun,\lambda}$ is the absolute magnitude of the Sun at wavelength $\lambda$, and $\Sigma$ is measured in ${M}_{\sun}$pc$^{-2}$. 
To evaluate the above expression, we need to obtain the $M/L$ ratio
at each radius. Following the prescription of Bell et al. (2003), we have
calculated the $M/L$ ratio as a function of color.

In this work we assume a Kroupa IMF (Kroupa 2001), which according to Bell et al. (2003) implies a deduction of 0.15 dex from the $M/L$ using the following expression:\\
\begin{equation}
 log_{10} (M/L)_{\lambda} = \left(a_{\lambda}\ +\ b_{\lambda}\ \times\ color \right)- 0.15, 
\end{equation}
where for $(g' - r')$ color, $a_{\lambda} = -0.306$ and $b_{\lambda} = 1.097$
is applied to determine the $r'$-band $M/L$. The resulting stellar mass density profiles are shown in the bottom row of Figure \ref{colorgrads}.

The Galactic extinction has been taken into account as described in PT06 using
the Schlegel et al. (1998) values:
\begin{equation}
 \mu_{corrected,\lambda} = \mu_{measured,\lambda} - A_{\lambda},
\end{equation}
where $A_{\lambda}$ is the extinction coefficient in each band.

\section{Results}

\subsection{Averaged Radial Surface Brightness}

The upper row of Figure \ref{colorgrads} shows the averaged radial surface
brightness profiles for the Type~I, Type~II and Type~III galaxies. The
increase towards the center over the inwards extrapolated (inner) disk
starting typically at around 0.2$R/R_b$ is due to the presence of bulges.
In the outer regions the characteristic break features, truncations and
antitruncations, are clearly seen in the mean profiles of Type~II and III
galaxies, respectively.

\subsection{Color Gradients}

The middle row of Figure \ref{colorgrads} shows the $(g' - r')$ radial color
profiles. It is interesting to note that each galaxy type has its own
characteristic color gradient. As found in previous works (e.g., de Jong
1996), the disks exhibit a general bluing as a function of their radius. This
is seen for all the galaxy types.

The color of Type~I galaxies, after reaching an asymptotic value of $(g' - r')
\sim 0.46$ mag in the outer regions ($\sim2R_h$), stays within the
error bars unchanged beyond. Type~II galaxies show a minimum [at $(g' - r') =
0.47 \pm 0.02$ mag] in their color profile at the break radius with the
profile getting redder beyond. After the initial bluing, the color of Type~III
galaxies gets redder towards the break radius to a mean value of $(g' - r') =
0.57 \pm 0.02$ mag. It is important to note that we can recover the
above color behavior of the mean profile basically for every individual galaxy
of each subsample, so consequently this is not an artifact of our profile
combination. In the outermost region of the profile, the uncertainty in the sky
determination can slightly increase the error bar on the color determination.
We have estimated this to be a factor of $\sim\sqrt{2}$ larger.

\begin{figure}
\centering
\includegraphics[scale=1.0]{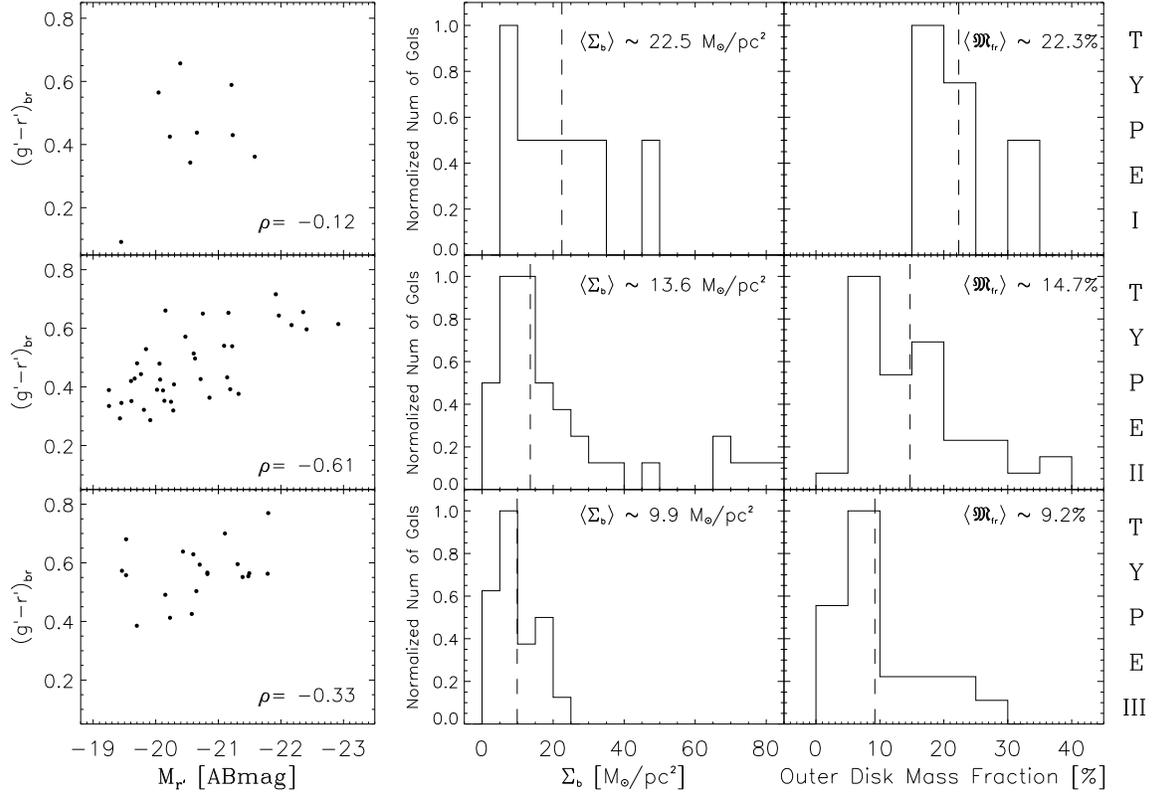}
\caption{\textit{Left column}: Absolute $r'$ magnitude and break color
  (or at 2.5 scale lengths in the case of Type~I galaxies) correlations; $\rho$ is
  the Spearman's correlation coefficient. Type~II galaxies show a strong
  correlation between the break color and the absolute magnitude, which means
  that the scatter in break color at a given luminosity is significantly
  smaller than the overall scatter. \textit{Middle and right columns}: Stellar surface
  mass density and stellar mass fraction histograms with their median values
  in the upper right corner of each panel.}
\label{hists}
\end{figure}

All galaxy types have a feature in common (see Fig.~\ref{colorgrads}): a large
scatter of the surface brightness and color profiles. To understand the origin
of this scatter, we have explored how the color at the break position
correlates with different properties of the galaxies. We find that the scatter
is best correlated with the total absolute magnitude of the individual
galaxies (see Fig.~\ref{hists}, left column).
To quantify the strength of these correlations, we have performed Spearman
correlation analysis. For all the three types we find that the brighter the
galaxy, the redder the color at the break. This correlation is weaker for
Type~I galaxies, because the sample is too small to provide reliable
statistics, but becomes particularly clear for Type~II profiles where our
$r'$-band absolute magnitude range is the largest. Figure \ref{hists} shows that
at a fixed absolute magnitude, the range of the $(g' - r')$ color at the break
is only $\sim$ 0.15 mag. This is a factor of 2 smaller than the
overall range in Figure \ref{colorgrads}.

\subsection{Surface Mass Density Profiles}

As explained in \S \ \ref{data}, we have obtained $M/L$ ratio
profiles from the $(g' - r')$ colors. These profiles were then converted into
stellar surface mass density profiles, which we will discuss here. The most
striking result is that both Type~I and Type~II galaxy profiles look very
similar, even without any quantitative measurement of the steepness. The
break for the Type~II galaxies that is so apparent in the light profiles has almost (for some individual galaxies completely) disappeared. In case of Type~III galaxies, the shape of the profile has not changed dramatically: a change of the slope around the break is still evident;
however, the slope becomes less well described by two individual exponentials.

\subsection{Break Surface Mass Density and Stellar Mass Fraction beyond the Break Radius}

Since the stellar surface mass density is an important tracer of star
formation and disk stability (Kennicutt 1989), we provide here numbers
corresponding to the break position.
The middle column of Figure \ref{hists} shows the histograms of stellar surface
mass density at the break radius or, for Type~I galaxies, at $R\!=\!2.5R_{h}$. The values of $\Sigma_{br}$ are $22.5\ \pm\ 5.3\ {M}_{\sun}$pc$^{-2}$ (Type~I), $13.6\ \pm\ 1.6\ {M}_{\sun}$pc$^{-2}$ (Type~II), and $ 9.9\ \pm\ 1.3\ {M}_{\sun}$pc$^{-2}$ (Type~III). Note that we do not detect any brek on the stellar mass density profile of Type I galaxies down to $\sim3 \ {M}_{\sun}$pc$^{-2}$

Another quantity we have calculated is the amount of stellar mass beyond the
break radii, which can be very useful for constraining the theoretical model. In
the right column of Figure \ref{hists}, the stellar mass fractions in the outer disk are
shown. Type~I galaxies contain about $22.3\ \pm \ 2 \%$ of their total stellar
mass beyond a radius of 2.5 times their scale-length. Type~II and Type~III
galaxies have much less stellar mass in the outer regions. For Type~III
galaxies ($9.2\ \pm\ 1.4 \% $) the amount of stellar mass is the lowest, 
with Type~II in between ($14.7\ \pm\ 1.2 \%$).

\section{Discussion}

How can the results found in this work be used to constrain the current models for the
formation of breaks in the surface brightness profiles of disk galaxies?

In the case of Type~II galaxies, the traditional pictures of
break formation (see \S \ \ref{intro}) grouped into two families: angular
momentum versus thresholds of star formation. Neither of these two ideas taken at
face value can explain why we find so many stars beyond the break radius, so
we will not go into more detail on these individual models. However, it is
important to stress that the new rendition of models has been able to
naturally explain the existence of stars beyond the break radius, and even
more, the exponential nature of the shape of the surface brightness beyond
this feature.
In particular, in the case of the Ro\v skar et al. (2008) model in which the
breaks are the result of the interplay between a radial star formation cutoff
and a redistribution of stellar mass by secular processes, a natural prediction
is the existence of a minimum in the age of the stellar population at the
break position, and a further aging (and consequently, a likely reddening) of
these stars as we move farther and farther away from the break radius. This is
in qualitative agreement with what we see in our color profiles for this
kind of galaxy. However, what these models have not been able to reproduce is
the absence of a clear break in the stellar mass density profile. 

The near absence of a break in the stellar surface mass density profile for our galaxies gives a strong indication that the behavior of the surface brightness profile outside the break is basically due to a change in the ingredients of the stellar population. If the shape of the color we see is not caused by a change in metallicity, this behavior could be explained as a natural consequence of stochastic migration of young stars from the inner parts of the disk to the outskirts (Ro\v skar et al. 2008). This will result in an age gradient where the oldest stars are the dominant component in the outskirts of the disks.

Unfortunately, other models (like Bournaud et al. 2007 and Foyle et al. 2008)
that are capable of creating stellar mass density profiles that resemble the
Type~II one we have found here (i.e., without a clear break) do not provide
any prediction on the color (age) distribution the stars should have along
the radial range.

Nevertheless, the fact that the stellar mass density of the break for this
type of galaxy with $\sim$13 ${M}_{\sun}$pc$^{-2}$ is so close to the
gas density threshold prediction of $\sim$10 ${M}_{\sun}$pc$^{-2}$
makes the case for a stellar population origin for the surface brightness
break even stronger (with a 100\% efficiency of transforming gas to stars).

Evidence for the same color phenomenology for Type~II galaxies also at high
redshift is presented by Azzollini et al. (2008). They have shown that a
similar minimum in the color profile can be found, at least, up to z $\sim$ 1
and that the main source of the scatter of the color profiles is caused by
the different stellar mass (in our case, absolute magnitude) of the galaxies in
their sample. Following these findings, we can conclude that once the absolute
magnitude of a galaxy is fixed, the color profiles within a given Type~(I, II or
III) of galaxy are strikingly similar. A more massive galaxy has a redder global color but the same shape of the color gradient.

Combining the results found in Azzollini et al. (2008) with what we find here,
one is tempted to claim that both the existence of the break in Type~II
galaxies and the shape of their color profiles are long-lived
features in the galaxy evolution, because it would be hard to imagine
how the above features could be continuously destroyed and recreated
maintaining the same properties over the last $\sim$8 Gyr.

In the case of Type~III galaxies, the situation is less clear. On the one hand,
our sample is smaller than in the case of Type~II galaxies and consequently our results
less robust. And on the other hand, the theoretical models are less elaborated
than for truncated galaxies, and no clear predictions have been made in
particular for the color profiles. Taking into account that the shape of the
stellar mass density profile does not differ too much from what we see in the
surface brightness profiles, we are inclined to think that our data do not
favor a sole origin in stellar population changes for this type of galaxy
but an authentic change in the amount of stars from the exponential
continuation of their inner region. It is interesting to note that in most of
the proposed ideas summarized in \S \ \ref{intro} to explain this kind
of galaxy, the origin of the stars in the periphery are linked to a dynamical
(in some case external) origin. So what we are seeing for these galaxies is
maybe a combination of star formation combined with an infall of new stars
from a external (satellite) source.

To corroborate our results, we plan to increase the number of galaxies as
well as the number of observed filters in our next study.

\acknowledgments
This research was supported by the Instituto Astrof\'isica de Canarias.
We thank Alexande Vazdekis and Ruym\'an Azzollini for their valuable comments and the anonymous referee
for his or her careful reading.


\clearpage

\end{document}